\documentclass[12pt]{elsarticle}

\usepackage{xcolor}
\usepackage{pdfsync}
\usepackage{amsmath}
\usepackage{amsfonts}
\usepackage{amssymb}
\usepackage{graphicx}
\usepackage{bbm}
\usepackage{multirow}
\usepackage{amssymb}
\usepackage{multicol}
\usepackage{booktabs}
\usepackage{epsfig}
\usepackage{subfig}
\usepackage[english]{babel}
\usepackage{lscape}

\usepackage{eurosym}
\newtheorem{teo}{Theorem}[section]

\newproof{pf}{Proof}

\def \R  {{\mathbb {R}}}
\begin{document}

\title{PDE models for the valuation of a non callable defaultable coupon bond under an extended JDCEV model}

\author[udc]{M.C. Calvo-Garrido}
\ead{mcalvog@udc.es}

\author[unibo]{S. Diop}
\ead{sidy.diop2@unibo.it}

\author[unibo]{A. Pascucci}
\ead{andrea.pascucci@unibo.it}

\author[udc]{C. V\'azquez \corref{cor1}}
\ead{carlosv@udc.es}
\cortext[cor1]{Corresponding author}
\address[udc]  {Department of Mathematics and CITIC, University of A Coru\~na. Campus Elvi\~na s/n, 15071 A Coru\~na, Spain}
\address[unibo] {Department of Mathematics, University of Bologna, 40126 Bologna, Italy}

\begin{abstract}
We consider a two-factor model for the valuation of a non callable defaultable bond which pays
coupons at certain given dates. The model under consideration is the Jump to Default Constant
Elasticity of Variance (JDCEV) model. The JDCEV model is an improvement of the
reduced form approach, which unifies credit and equity models into a single framework allowing for
stochastic and possible negative interest rates.
From the mathematical point of view, the
valuation involves two partial differential equation (PDE) problems for each coupon. First, we
obtain the existence of solution for these PDE problems. In order to solve them, we propose
appropriate numerical schemes based on a Crank-Nicolson semi-Lagrangian method for time
discretization combined with biquadratic Lagrange finite elements for space discretization. Once
the numerical solutions of the PDEs are obtained, a post-processing procedure is carried out in
order to achieve the value of the bond. This post-processing includes the computation of an
integral term which is approximated by using the composite trapezoidal rule. Finally, we present
some numerical results for real market bonds issued by different firms in order to illustrate the
proper behaviour of the numerical schemes. Moreover, we obtain an agreement between the numerical
results from the PDE approach and those ones obtained by applying a Monte Carlo technique and an
asymptotic aproximation method.
\end{abstract}
\maketitle
\section{Introduction}
Default risk can be modelled by using two different approaches: the structural approach and the reduced-form approach. For more information about the history, advantages and drawbacks of each one of the two approaches we refer the readers to \cite{agliardi2011, ce2003, ce2006}. In the structural method default occurs when the firm asset value reaches some lower barrier, whereas in the reduced-form approach, the default event is assumed to be unpredictable and governed by a default intensity process that could be either deterministic or stochastic. Moreover, in the reduced-form approach the default event can occur without any correlation with the firm value. In this work we consider the reduced-form approach.

In the literature there are several papers devoted to default risk modelling by using the two different approaches. On one hand, the first approach is taking into consideration in \cite{agliardi2011} where the author proves an exact formula for the valuation of defaultable coupon bonds by generalizing the one derived in \cite{geske1977}. On the other hand, both approaches are combined in \cite{ce2006} also to price defaultable bonds.

The main objective of this paper is to obtain the price of a defaultable coupon bond under the extended Jump to Default Constant Elasticity of Variance (JDCEV) model proposed in \cite{cl2006}, also considering the possibility of incorporating negative interest rates which is introduced in \cite{ddp2017}. The JDCEV model was first introduced in \cite{cl2006} as a hybrid credit and equity model, although only taking into account constant positive interest rates. In this work we assume that the interest rates dynamics is governed by a stochastic process which takes into account the possibility of negative interest rates as in \cite{ddp2017}. The incorporation of negative interest rates results more realistic according to the current situation of real markets. As it was pointed out before, this model can be set in the framework of the reduced-form approaches. More precisely, in the present paper we define the instantaneous volatility of the stock price as a constant elasticity of variance (CEV) process and we assume that the default intensity is an affine function of the instantaneous variance of the underlying stock.

From the mathematical point of view, the valuation problem of a defaultable coupon bond can be posed in terms of a sequence of partial differential equation (PDE) problems, where the underlying stochastic factors are the interest rates and the stock price. Moreover, the stock price follows a diffusion process interrupted by a possible jump to zero (default), as it is indicated in \cite{cl2006}. In order to compute the value of the bond we need to solve two partial differential equation problems for each coupon, with maturities equal to those coupon payment dates. Concerning the numerical solution of these PDE problems, after a localization procedure to formulate the problems in a bounded domain and the study of the boundaries where boundary conditions are required following the ideas introduced in \cite{olerad1973}, we propose appropriate numerical schemes based on a Crank-Nicolson semi-Lagrangian method for time discretization combined with biquadratic Lagrange finite elements for space discretization. The numerical analysis of this Lagrange-Galerkin method has been addressed in \cite{bnv2006,bnv2006b}. Once the numerical solution of the PDEs is obtained, a kind of post-processing is carried out in order to achieve the value of the bond. This post-processing includes the computation of an integral term which is approximated by using the composite trapezoidal rule.

This paper is organized as follows. In Section \ref{model1}, we describe the  two stochastic factors (i.e. the interest rate and the defaultable stock price) that are involved in the model, and we state the PDE problem that governs the valuation of non callable defaultable coupon bonds. In Section \ref{existencesolution} we establish the existence of solution for the PDE problems. In Section \ref{numericalmethod}, we formulate the pricing problem in a bounded domain after a localization procedure and we impose appropriate boundary conditions. Then, we introduce the discretization in time of the problem by using a Crank-Nicolson characteristic scheme, and we state the variational formulation of the problem in order to apply finite elements for the discretization in the asset and interest rate variables. In Section \ref{results} we present some numerical results to illustrate the good performance of the models and numerical methods, also including a comparison with the results obtained with an alternative Monte Carlo technique and an asymptotic expansion method. Finally, we finish with some conclusions in Section \ref{conclusions}.

\section{Mathematical modelling}\label{model1}
\subsection{Stochastic underlying variables}
A non callable defaultable coupon bond is a financial derivative product, the underlying variables of which are the interest rate and the defaultable stock price.
The interest rate at time $t$, $r_t$, is assumed to be stochastic and its dynamics under a risk neutral probability measure is driven by the Vasicek model \cite{v1977}, in which the spot rate is governed by the Ornstein-Uhlenbeck process:
\begin{equation}
dr_t \, = \, \kappa(\theta - r_t) \, dt \, + \, \delta \, dW_t^1,
\end{equation}
where $\kappa>0$ is the speed of adjustment in the mean reverting process, $\theta>0$ is the long-term mean of the short-term interest rate, $\delta>0$ is the interest rate volatility and
$W_t^1$ is the standarized Wiener process for the interest rate. Negative interest rates are taken into consideration with this model contrary to other interest rate dynamics, such as CIR model \cite{cir1985}. Note that the possibility of negative interest rates results more realistic according to the current situation of the markets.

In the JDCEV model we assume that the local volatility of the stock is given by
\begin{displaymath}
\sigma(t,S_t)=a(t)\,S_t^{\beta},
\end{displaymath}
where $S_t$ denotes the stock price at time $t$, $\beta<0$ is the elasticity parameter and $a(t)>0$ is the time dependent volatility scale function. Moreover, the default intensity can be written in terms of the stock volatility and the stock price in the following way
\begin{equation}\label{lambda}
\lambda(t,S_t)=b(t)+c\,\sigma(t,S_t)^2=b(t)+c\,a(t)^2\,S_t^{2\beta},
\end{equation}
where $b(t)\geq0$ is a deterministic non-negative function of time and $c>0$ governs the sensitivity of the default intensity with respect to the volatility, as it is indicated in \cite{cl2006}.
The other source of uncertainty, the predefaultable stock price at time t, $S_t$, under a risk neutral probability measure satisfies the following stochastic differential equation:
\begin{equation}\label{stockprice}
dS_t \, = \, (r_t+\lambda(t,S_t))S_t \, dt\, + \, \sigma(t,S_t)\,S_t\,dW_t^2.
\end{equation}
where $W_t^2$ is the standarized Wiener process for the stock price.
Both Wiener processes, $W_t^1$ and $W_t^2$ can be correlated according to $dW_t^1 dW_t^2=\rho dt$, where $\rho$ denotes the instantaneous correlation coefficient such that $\mid \rho \mid < 1$. Moreover, in this model the time of default $\xi$ has two parts. On one hand, we can define the predictable part as $\xi_0=\text{inf}\{t\geq0: S_t=0\}$, whereas on the other hand the random part $\tilde{\xi}$ is given by the following expresion:
\begin{equation}
\tilde{\xi}=\text{inf}\left\{t \geq 0:\int_0^t \lambda(t,S_t)\geq e\right\},
\end{equation}
where $e$ is an exponential random variable $e \sim Exp(1)$. Thus, the time of default is defined as $\xi=\xi_0 \wedge \tilde{\xi}$.

Finally, the defautable stock price is given by $\overline{S}_t=S_t{\mathbbm{1}}_{ \{ \xi >t \}}$.
\subsection{PDE formulation}\label{pdeformulation}
In this work we consider the valuation of a non callable defaultable bond with the possibility of paying a series of coupons at given dates $t_i$, for $i=1,..., M$, where $M$ is the number of coupons and $T=t_M$ is the maturity of the bond. At maturity, the bond holder receives the face value (FV) plus the last coupon. Thus, let us denote by $cp_i$ the amount of money paid at coupon payment date $t_i$. This amount is computed by multiplying the coupon rate, the frequency of coupon payments and the FV. Having this in view, the value of a non callable defaultable coupon bearing bond at time $t=t_0=0$ for the spot values $S_0$ and $r_0$, $V(0,S_0,r_0;T)$, which is understood as the discounted value of the future coupon payments and the face value of the bond, is given by
{\small
\begin{eqnarray}\label{bondValueExpectations}
&&V(0,S_0,r_0;T)=FV\left[\sum_{i=1}^M cp_i\,\mathbb{E}\left[\exp\left( -\int_0^{t_i}\left( r_u+\lambda(u,S_u)\right)du\right)\right]+\mathbb{E}\left[\exp\left(-\int_0^{T}\left( r_u+\lambda(u,S_u)\right)du\right)\right] \right.\nonumber\\
&&+\eta \left.\left(1-\mathbb{E}\left[\exp\left( -\int_0^{T}\left( r_u+\lambda(u,S_u)\right)du\right)\right] -\int_0^T \mathbb{E}\left[\exp\left(-\int_0^{\tau_1}\left( r_u+\lambda(u,S_u)\right)du\right)r_{\tau_1}\right]\,d \tau_1\right)\right],
\end{eqnarray}
}

where $\eta$ is the recovery rate in case of default. An analysis of the recovery rate by industrial sector and debt seniority is carried out in \cite{ak1996}.

Next, if we denote by
\begin{eqnarray*}
u_1(0,S_0,r_0;t_i)&=&\mathbb{E}\left[\exp\left( -\int_0^{t_i}\left( r_u+\lambda(u,S_u)\right)du\right)\right],\\
u_1(0,S_0,r_0;T)&=&\mathbb{E}\left[\exp\left( -\int_0^{T}\left( r_u+\lambda(u,S_u)\right)du\right)\right],\\
u_2(0,S_0,r_0;\tau_1)&=&\mathbb{E}\left[\exp\left(-\int_0^{\tau_1}\left( r_u+\lambda(u,S_u)\right)du\right)r_{\tau_1}\right],
\end{eqnarray*}
then the expression of the bond value (\ref{bondValueExpectations}) can be written equivalently as
\begin{eqnarray}\label{bondValue}
V(0,S_0,r_0;T)&=&FV\left[\sum_{i=1}^M cp_i\, u_1(0,S_0,r_0;t_i)+u_1(0,S_0,r_0;T)\right.\nonumber\\&&\left.+\eta \left(1-u_1(0,S_0,r_0;T)-\int_0^T u_2(0,S_0,r_0;\tau_1)\,d \tau_1 \right)\right].
\end{eqnarray}

Moreover, by applying the Feynman-Kac formula (see \cite{pascucci2011}, for example) and using the change of variable $y_t=r_t\exp(\kappa t)$, the functions $u_1$ and $u_2$ are the solutions of the Cauchy problem
\begin{equation}\label{cauchyproblem}
\left\{ \begin{array}{ll}
\overline{\mathcal{L}}\left[u(t,S,y)\right]=0, & t<T_1, \, (S,y) \in (0,\,\infty)\times (-\infty,\,\infty),\\
u(T_1,S,y)=h(S,y), & (S,y) \in (0,\,\infty)\times (-\infty,\,\infty),
\end{array}\right.
\end{equation}
with $h(S,y)=1$ for $u=u_1$ or $h(S,y)=\exp(-\kappa T_1)y$ for $u=u_2$, respectively. Moreover, the operator $\overline{\mathcal{L}}$ is defined as follows
\begin{eqnarray}\label{operator}
\overline{\mathcal{L}}[u] & = & \partial_t u+\frac{1}{2}\sigma^2(t,S) S^2 \, \partial_{SS} u+\rho \delta \sigma(t,S)\exp(\kappa t) S\, \partial_{Sy} u +\frac{1}{2} \delta^2 \exp(2 \kappa t)\, \partial_{yy} u\nonumber\\ &&
+\left(\exp(-\kappa t) y + \lambda(t,S)\right) S\, \partial_S u+ \kappa \theta \exp(\kappa t) \,\partial_y u-\left(\exp(-\kappa t) y + \lambda(t,S)\right)u.
\end{eqnarray}

\section{Existence of solution}\label{existencesolution}
In this section, we aim to prove the existence of a ``local'' solution to the Cauchy problem
\eqref{cauchyproblem}. For this purpose, let $X$ be the logarithm of the pre-defaultable stock
price, i.e. $X_t = \log{S_t}$, $t\in[0,T]$. Our model becomes
\begin{equation}\label{model}
\left\{ \begin{array}{ll}
\overline{S}_t = S_0 \exp(X_t){\mathbbm{1}}_{ \{ \xi >t \}} \qquad S_0 > 0 \\
d r_t = \kappa(\theta- r_t) d t +  \delta \d W_t^1 \\
d X_t = (r_t-\frac{1}{2}\sigma(t,X_t)^2 + \lambda(t, X_t)) d t + \sigma(t, X_t)d W_t^2\\
dW^1_t dW^2_t = \rho dt,
\end{array}\right.
\end{equation}
with $\sigma(t, x) = a(t)e^{\beta x}$ and $\lambda(t,x) = b(t)+c\sigma(t,x)^2$. The corresponding
infinitesimal generator is the operator $\mathcal{L}$ defined as
\begin{eqnarray}\
\mathcal{L} & = & \partial_t +\frac{1}{2}\sigma^2(t,x) \, \partial_{xx}+\rho \delta \sigma(t,x)\, \partial_{xr}  +\frac{1}{2} \delta^2 \, \partial_{rr} \nonumber\\
 & & +\left(r -\frac{1}{2}\sigma(t,x)+ \lambda(t,x)\right) \, \partial_x + \kappa\left(\theta-r\right) \,\partial_r -\left(r +
 \lambda(t,x)\right)\\
 & = & \partial_t +\frac{1}{2}\langle\Sigma \nabla,\nabla\rangle+\langle \mu,\nabla\rangle+\gamma
\end{eqnarray}
where
\begin{eqnarray}
 \Sigma\left(t, x, r\right) & = & \left(\begin{array}{cc} \sigma^2\left(t,x\right) & \rho\delta\sigma\left(t,x\right)\\ \rho\delta\sigma\left(t,x\right) & \delta^2 \end{array}\right), \quad \mu\left(t,x,r\right)  = \left(\begin{array}{c} r-\frac{1}{2}\sigma^2\left(t,x\right)+\lambda\left(t,x\right) \\ \kappa\left(\theta-r\right)\end{array}\right),\nonumber\\ \gamma\left(t,x\right)  & = & -\left(r+\lambda\left(t,x\right)\right).
\end{eqnarray}
Operator $\mathcal{L}$ is {\it only locally} uniformly parabolic in the sense that, for any ball
 $$\mathcal{O}_{R}:=\left\{(x,r)\in\R^{2}\mid\left|(x,r)\right|< R\right\};$$
the coefficients of $\mathcal{L}$ satisfy the following conditions: 

\textbf{(H1)} The matrix $\Sigma(t,x,r)$ is positive definite,  uniformly with respect to
$(t,x,r)\in \left(0,T\right]\times \mathcal{O}_{R}$.

\textbf{(H2)} The coefficients $\Sigma,\mu,\gamma$ are bounded and H\"older-continuous on
$\left(0,T\right]\times\mathcal{O}_{R}$.

Under these conditions we can resort on the recent results in \cite{pp2017}, Theor. 2.6, or
\cite{lpp2018}, Theor.1.5, about the existence of a {\it local density} for the process
$(X,r)$.
\begin{teo}
For any $R>0$, the process $(X,r)$ has a local transition density on $\mathcal{O}_{R}$, that is a
non-negative measurable function $\Gamma=\Gamma(t,x,r;T,z,s)$ defined for any $0<t<T$,
$(x,r)\in\R^{2}$ and $(z,s)\in\mathcal{O}_{R}$ such that, for any continuous function $h=h(x,r)$
with compact support in $\mathcal{O}_{R}$, we have
  $$u(t,x,r):=\mathbb{E}\left[h(X_{T},r_{T})\right]=\int_{\mathcal{O}_{R}}\Gamma(t,x,r;T,z,s)h(z,s)dz ds$$
and $u$ satisfies
\begin{equation}\label{and1}
  \begin{cases}
    \mathcal{L}u(t,x,r)=0, & (t,x,r)\in[0,T)\times\mathcal{O}_{R}, \\
    u(T,x,r)=h(x,r) & (x,r)\in \mathcal{O}_{R}.
  \end{cases}
\end{equation}
\end{teo}

Problem \eqref{and1} can be used for numerical approximation purposes. However, notice that
\eqref{and1} does not have a unique solution due to the lack of lateral boundary conditions.
Nevertheless, numerical schemes can be implemented imposing artificial boundary conditions and the
result which guarantees the validity of such approximations is the so-called {\it principle of not
feeling the boundary}. A rigorous statement of this result can be found in \cite{gatheraletall2012},
Appendix A, or in \cite{pp2017}, Lemma 4.11.

\section{Numerical methods}\label{numericalmethod}
In order to obtain a numerical approach of the value of a non callable defaultable coupon bond we need to solve the Cauchy problem (\ref{cauchyproblem}) for $u=u_1$ and $u=u_2$ with maturity $T_1=t_i$ for $i=1,..., M$, that is each coupon payment date for both cases. Once these problems are solved, the value of the bond is given by expression (\ref{bondValue}) in which an integral term appears. This integral term will be approximated by means of the classical composite trapezoidal rule. For the numerical solution of the PDE problem, we propose a Crank-Nicolson characteristics time discretization scheme combined with a piecewise bi-quadratic Lagrange finite element method. The convergence properties of this Lagrange-Galerkin method have been mathematically analyzed in \cite{bnv2006,bnv2006b} for time and space discretization. More recently, it has been applied to the valuation of pension plans without and with early retirement in \cite{cv2012} and \cite{cpv2013}, respectively. In order to apply this set of numerical techniques, first a localization procedure is used to cope with the initial formulation in an unbounded domain.

\subsection{Localization procedure and formulation in a bounded domain}
In this section we replace the unbounded domain by a bounded one. In order to determine the required boundary conditions for the associated PDE problem we follow \cite{olerad1973} which is based on the theory proposed by Fichera in \cite{fichera1960}.
Let us introduce the following notation:
\begin{equation}
x_0=t, \quad \tilde{x}_1=S \quad \text{and} \quad \tilde{x}_2=y.
\end{equation}
For this purpose, let us consider both $\tilde{x}_1^\infty$ and $\tilde{x}_2^\infty$ to be large enough suitably chosen real numbers such that the solution in the region of financial interest is not affected by the truncation of the domain.
Let
\begin{displaymath}
\Omega^*=(0,x_0^\infty)\times\left(\frac{1}{\tilde{x}_1^\infty},\tilde{x}_1^\infty\right)\times(-\tilde{x}_2^\infty ,\tilde{x}_2^\infty)
\end{displaymath}
with $x_0^\infty=T_1$.
Additionally, we make the changes of variables $$x_1=\tilde{x}_1-\frac{1}{\tilde{x}_1^\infty}, \quad x_2=\tilde{x}_2+\tilde{x}_2^\infty \, ,$$ as well as the values $x_1^\infty=\tilde{x}_1^\infty-\frac{1}{\tilde{x}_1^\infty}$ and $x_2^\infty=2\tilde{x}_2^\infty$,
which leads to replace the bounded domain $\Omega^*$ by the following one:
\begin{displaymath}
\tilde{\Omega}=(0,x_0^\infty)\times(0,x_1^\infty)\times(0,x_2^\infty)
\end{displaymath}
Then, let us denote the Lipschitz boundary by $\tilde{\Gamma}=\partial \tilde{\Omega}$ such that $\tilde{\Gamma}=\bigcup_{i=0}^2(\tilde{\Gamma}_i^{-}\cup\tilde{\Gamma}_i^{+})$, where
$$\tilde{\Gamma}_i^{-}=\{(x_0,x_1,x_2)\in\tilde{\Gamma}\mid x_i=0\}, \quad
\tilde{\Gamma}_i^{+}=\{(x_0,x_1,x_2)\in\tilde{\Gamma} \mid x_i=x_i^\infty\}, \quad i=0, 1, 2.$$
Then, the operator defined in (\ref{operator}) can be written in the form:
\begin{equation}\label{pdeOleinik}
\mathcal{L}[u]=\sum_{i,j=0}^2b_{ij}\frac{\partial^2u}{\partial x_ix_j}+\sum_{j=0}^2b_j\frac{\partial u}{\partial x_j}+b_0u,
\end{equation}
where the involved data are given by
{\small
\begin{eqnarray}
\mathbf{B}&=&(b_{ij})=\left(\begin{array}{ccc}0 & 0 & 0\\
0 & \frac{1}{2}a^2(t)\left( x_1+\frac{1}{\tilde{x}_1^\infty} \right)^{2\beta+2} & \frac{1}{2}\rho\delta a(t) \left(x_1+\frac{1}{\tilde{x}_1^\infty}\right)^{\beta+1}\exp(\kappa t) \\
0 & \frac{1}{2}\rho\delta a(t)\left(x_1+\frac{1}{\tilde{x}_1^\infty} \right)^{\beta+1}\exp(\kappa t) & \frac{1}{2}\delta^2 \exp(2\kappa t)
 \end{array}\right) , \\ \nonumber\\
 \mathbf{b}&=&(b_j)=\left(\begin{matrix}
 1\\
\left(\exp(-\kappa t)(x_2-\tilde{x}_2^\infty)+\lambda\left(t,x_1+\frac{1}{\tilde{x}_1^\infty}\right)\right)\left(x_1+\frac{1}{\tilde{x}_1^\infty}\right)\\
\kappa\theta\exp(\kappa t)
\end{matrix}\right), \\ \nonumber\\
b_0&=&-\left(\exp(-\kappa t)(x_2-\tilde{x}_2^\infty)+\lambda\left(t,x_1+\frac{1}{\tilde{x}_1^\infty}\right)\right).
\end{eqnarray}
}

Thus, following \cite{olerad1973}, in terms of the normal vector to the boundary pointing inward $\tilde{\Omega}$, $\mathbf{m}=(m_0,m_1,m_2)$, we introduce the following subsets of $\tilde{\Gamma}$:
\begin{equation}
\Sigma^0=\left\{x\in\tilde{\Gamma}/\sum_{i,j=0}^2b_{ij}m_im_j=0\right\},\quad
\Sigma^1=\tilde{\Gamma}-\Sigma^0, \nonumber
\end{equation}
\begin{equation}
\Sigma^2=\left\{x\in\Sigma^0/\sum_{i=0}^2\left(b_i-\sum_{j=0}^2\frac{\partial b_{ij}}{\partial x_j}\right)m_i<0\right\}. \nonumber
\end{equation}
As indicated in \cite{olerad1973} the boundary conditions at $\Sigma^1\bigcup\Sigma^2$ for the so-called first boundary value problem associated with (\ref{pdeOleinik}) are required.
Note that $\Sigma^1=\{\tilde{\Gamma}_1^{-},\, \tilde{\Gamma}_1^{+},\, \tilde{\Gamma}_2^{-},\, \tilde{\Gamma}_2^{+}\}$ and $\Sigma^2=\{\tilde{\Gamma}_0^{+}\}$. Therefore, in addition to a final condition (see section \ref{pdeformulation}), we need to impose boundary conditions on $\tilde{\Gamma}_1^{-}$, $\tilde{\Gamma}_1^{+}$, $\tilde{\Gamma}_2^{-}$ and $\tilde{\Gamma}_2^{+}$. Next, we will impose Dirichlet conditions on $\tilde{\Gamma}_1^-$, $\tilde{\Gamma}_2^-$ and $\tilde{\Gamma}_2^+$, whereas on $\tilde{\Gamma}_1^+$ we will impose a homogeneous Neumann condition.

Taking into account the previous change of spatial variable and making the change of time variable $\tau=T_1-t$, we write the equation (\ref{cauchyproblem}) in divergence form in the bounded spatial domain $\Omega=(0,x_1^\infty)\times(0,x_2^\infty)$. Moreover, we decompose the boundary $\partial \Omega=\bigcup_{i=1}^2({\Gamma}_i^{-}\cup\Gamma_i^{+})$ of the spatial domain $\Omega$ as follows:
$$\Gamma_i^{-}=\{(x_1,x_2)\in\Gamma\mid x_i=0\}, \quad
\Gamma_i^{+}=\{(x_1,x_2)\in\Gamma \mid x_i=x_i^\infty\}, \quad i=1, 2.$$

Thus, the initial boundary value problem (IBVP) takes the following form:

Find $u : [0,T_1] \times \Omega\rightarrow\mathbb{R}$ such that
\begin{eqnarray}\label{ivp}
\partial_\tau u-Div
(\mathbf{A}\nabla u )+\mathbf{v}\cdot\nabla u+l u=0 & &\mbox{in } (0,T_1) \times \Omega \, , \\
 u(0,.)=g &  & \hbox{in } \Omega,\\
u=f & & \mbox{on  } (0,T_1) \times \left(\Gamma_1^- \cup \Gamma_2^- \cup \Gamma_2^+ \right),\\
\frac{\partial u}{\partial{x_1}}=0 & & \mbox{on  }  (0,T_1) \times \Gamma_1^+. \label{ivp1}
\end{eqnarray}
For problem (\ref{ivp})-(\ref{ivp1}), the diffusion matrix $\mathbf{A}$ and the velocity field $\mathbf{v}$ are given by

{\tiny
\begin{eqnarray*}
\mathbf{A} & = & \left( \begin{array}{lr}\label{divAvf1}
            \frac{1}{2}a^2(T_1-\tau) \left(x_1+\frac{1}{\tilde{x}_1^\infty}\right)^{2\beta+2} & \frac{1}{2}\rho \delta a(T_1-\tau) \left(x_1+\frac{1}{\tilde{x}_1^\infty} \right)^{\beta+1} \exp(\kappa(T_1-\tau))\\
   \frac{1}{2}\rho \delta a(T_1-\tau) \left(x_1+\frac{1}{\tilde{x}_1^\infty} \right)^{\beta+1} \exp(\kappa(T_1-\tau))   &  \frac{1}{2}\delta^2 \exp(2\kappa(T_1-\tau))
           \end{array}
    \right), \quad \\ \\\\
\mathbf{v}& =&  \left( \begin{array}{c}
         \frac{1}{2}a^2(T_1-\tau)(2\beta+2)\left(x_1+\frac{1}{\tilde{x}_1^\infty} \right)^{2\beta+1}-\left(\exp(-\kappa(T_1-\tau))\left(x_2-\tilde{x}_2^\infty\right)+\lambda\left(T_1-\tau,x_1+\frac{1}{\tilde{x}_1^\infty}\right)\right)\left(x_1+\frac{1}{\tilde{x}_1^\infty}\right)  \\
      \frac{1}{2}\rho \delta a(T_1-\tau)(\beta+1) \left(x_1+\frac{1}{\tilde{x}_1^\infty}\right)^{\beta} \exp(\kappa(T_1-\tau))-\kappa \theta\exp(\kappa(T_1-\tau))
           \end{array}
    \right),\quad
 \end{eqnarray*}
}

while the reaction function $l$, the initial condition $g$ and the function $f$ for the Dirichlet boundary conditions are defined as follows:
\begin{eqnarray*}
l(x_1,x_2) & =& \exp(-\kappa (T_1-\tau))(x_2-\tilde{x}_2^\infty)+\lambda\left(T_1-\tau,x_1+\frac{1}{\tilde{x}_1^\infty}\right),\\ \\
g(x_1,x_2) & = & h\left(x_1+\frac{1}{\tilde{x}_1^\infty},x_2-\tilde{x}_2^\infty \right) ,\\ \\
f(\tau, x_1,x_2) & = & \exp\left(-\int_{T_1-\tau}^{T_1}\left(\exp(-\kappa \tilde{u})(x_2-\tilde{x}_2^\infty)+\lambda\left(\tilde{u},x_1+\frac{1}{\tilde{x}_1^\infty}\right)\right)d\tilde{u}\right)g(x_1,x_2).
\end{eqnarray*}

\subsection{Time discretization}
The method of characteristics is based on a finite differences scheme for the discretization
of the material derivative, i.e., the time derivative along the characteristic
lines of the convective part of the equation (\ref{ivp}). The material derivative operator is given by $$\frac{D}{D \tau}=\partial_\tau + \mathbf{v}\cdot\nabla.$$

For a brief description of the method, we first define the characteristics curve through $\mathbf{x}=(x_{1},x_2)$ at time $\bar{\tau}$, $ X(\mathbf{x},\bar{\tau};s)$, which satisfies:

\begin{equation} \label{carac}
\frac{\partial}{\partial s} X(\mathbf{x},\bar{\tau};s) = \mathbf{v}(X(\mathbf{x},\bar{\tau};s)), \quad X(\mathbf{x},\bar{\tau};\bar{\tau})=\mathbf{x}.
\end{equation}

In order to discretize in time the material derivative in equation (\ref{ivp}), let us consider a number of time steps $N$, the time step $\Delta \tau=T/N$ and the time mesh points $\tau^n=n \Delta \tau,$  $n = 0,\frac{1}{2}, 1,\frac{3}{2}, \ldots,N$.

The material derivative approximation by the characteristics method for both problems is given by:
\begin{displaymath}
\frac{Du}{D \tau} \approx \frac{u^{n+1}-u^n\circ X^n}{\Delta \tau},
\end{displaymath}
where $u=u_1, u_2$ and $X^n(\mathbf{x})=X(\mathbf{x},\tau^{n+1};\tau^n)$. In this case, the solution of (\ref{carac}) is not computed analytically. Instead, we consider numerical ODE solvers to approximate the characteristics curves (see \cite{bnv2006}, for example). More precisely, in this work we employ the explicit second order Runge-Kutta method to approximate the values of $X^n(\mathbf{x})$.

Next, we consider a Crank-Nicolson scheme around $\left(X(\textbf{x},\tau^{n+1};\tau),\tau\right)$
for $\tau=\tau^{n+\frac{1}{2}}$. So, the time discretized equation for $u=u_1, u_2$ can be written as follows:\\

\hspace{0.5cm}Find $u^{n+1}$ such that:
\begin{eqnarray}\label{EDPcaract}
\frac{u^{n+1}(\mathbf{x})-u^n(X^{n}(\mathbf{x}))}{\Delta \tau}-\frac{1}{2}
Div(A\nabla u^{n+1})(\mathbf{x})-\frac{1}{2} Div(\mathbf{A} \nabla u^n)(X^{n}(\mathbf{x}))  & &  \nonumber\\
\hspace*{-1cm}+\frac{1}{2} (l \, u^{n+1}) (\mathbf{x})+ \frac{1}{2} (l \, u^n) (X^{n}(\mathbf{x})) & = & 0.
\end{eqnarray}

In order to obtain the variational formulation of the semi-discretized problem, we multiply (\ref{EDPcaract}) by a suitable test function, integrate in $\Omega$, use the classical Green formula and the non classical following one \cite{maria2005}:
\begin{eqnarray}\label{ncgreen}
\int_{\Omega}Div(\mathbf{A}\nabla u^n)(X^n(\mathbf{x}))\psi(\mathbf{x})d\mathbf{x} & = &
\int_{\Gamma}(\nabla X^n)^{-T}(\mathbf{x})\mathbf{n}(x)\cdot(\mathbf{A}\nabla u^n)(X^n(\mathbf{x}))\psi(\mathbf{x})dA_\mathbf{x} \nonumber\\
& &  \hspace*{-1cm} -\int_{\Omega}(\nabla X^n)^{-1}(\mathbf{x})(\mathbf{A}\nabla u^n)(X^n(\mathbf{x}))\cdot\nabla \psi(\mathbf{x})d\mathbf{x} \nonumber\\
& &  \hspace*{-1cm} - \int_\Omega Div((\nabla X^n)^{-T}(\mathbf{x}))\cdot(\mathbf{A}\nabla u^n)(X^n(\mathbf{x}))\psi(\mathbf{x})d\mathbf{x}.\nonumber\\
\end{eqnarray}
Note that, as the characteristics curves cannot be obtained analytically, the terms $(\nabla X^n)^{-1}(\mathbf{x})$ and $Div((\nabla X^n)^{-T}(\mathbf{x}))$ in (\ref{ncgreen}) are replaced by the following approximations (see \cite{bnv2006} for more details):

$$(\nabla X^n)^{-1}(\mathbf{x})=\mathbf{I}(\mathbf{x})+\Delta \tau \mathbf{L}^n(X^n(\mathbf{x}))+O(\Delta \tau^2),$$

$$Div((\nabla X^n)^{-T}(\mathbf{x}))=\Delta \tau\,\nabla\, Div\, (\mathbf{v}^n(X^n(\mathbf{x})))+O(\Delta \tau^2),$$

where $\mathbf{L}=\nabla \mathbf{v}$.

After the previous steps, we can write a variational formulation for the time discretized problem as follows:

Find $u^{n+1} \in H^1(\Omega)$ satisfying the Dirichlet boundary condition (21), such that:

\begin{eqnarray}
\frac{1}{\Delta \tau}\int_{\Omega} u^{n+1}(\mathbf{x})\psi(\mathbf{x})d\mathbf{x}+\frac{1}{2} \int_{\Omega} (\mathbf{A} \nabla u^{n+1})(\mathbf{x})\cdot\nabla\psi(\mathbf{x})d\mathbf{x}
+\frac{1}{2}\int_{\Omega} (lu^{n+1})(\mathbf{x})\psi(\mathbf{x})d\mathbf{x}&&\nonumber\\
=\frac{1}{\Delta \tau}\int_{\Omega}u^n(X^n(\mathbf{x}))\psi(\mathbf{x})d\mathbf{x}
-\frac{1}{2}\int_{\Omega}(\mathbf{A}\nabla u^n)(X^n(\mathbf{x}))\cdot \nabla \psi(\mathbf{x}) d\mathbf{x}&&\nonumber\\-\frac{\Delta \tau}{2}\int_{\Omega}\mathbf{L}^n(X^n(\mathbf{x}))(\mathbf{A}\nabla u^n)(X^n(\mathbf{x}))\cdot \nabla \psi(\mathbf{x}) d\mathbf{x}
-\frac{1}{2} \int_{\Omega}(lu^n)(X^n(\mathbf{x}))\psi(\mathbf{x})d\mathbf{x}
&&\nonumber\\-\frac{\Delta\tau}{2}
\int_\Omega \nabla\,Div\,(\mathbf{v}^n(X^n(\mathbf{x})))\cdot(\mathbf{A}\nabla u^n)(X^n(\mathbf{x}))\psi(\mathbf{x})d\mathbf{x}&&\nonumber\\
+\frac{1}{2}\int_{\Gamma}(\mathbf{I}(\mathbf{x})+\Delta \tau \mathbf{L}^n(X^n(\mathbf{x})))^T\mathbf{n}(x)\cdot(\mathbf{A}\nabla u^n)(X^n(\mathbf{x}))\psi(\mathbf{x})dA_\mathbf{x}+\frac{1}{2}\int_{\Gamma_1^+}a_{12}(\mathbf{x})\frac{\partial u}{\partial x_2}(\mathbf{x})\psi(\mathbf{x})dA_\mathbf{x},&&\nonumber\\
\end{eqnarray}
 for all $\psi\in H^1(\Omega)$ such that $\psi=0$ on $\Gamma_1^-$, $\Gamma_2^-$ and $\Gamma_2^+$, In the last term $a_{12}$ is the corresponding coefficient of the diffusion matrix $\mathbf{A}$.

\subsection{Finite elements discretization}
For the spatial discretization we consider $\{\tau_h\}$, a quadrangular mesh of the domain  $\Omega$. Let $(T_2,\mathcal{Q}_2,\Sigma_{T_2})$ be a family of piecewise quadratic Lagrangian finite elements, where $\mathcal{Q}_2$ denotes the space of polynomials defined in $T_2\in\tau_h$ with degree less or equal than two in each spatial variable and $\Sigma_{T_2}$ the subset of nodes of the element $T_2$. More precisely, let us define the finite elements space $u_h$ by
\begin{equation}
u_h=\{\phi_h\in\mathcal{C}^0(\Omega):\phi_{h_{T_2}}\in\mathcal{Q}_2,\quad\forall T_2\in\tau_h\},
\end{equation}
where $\mathcal{C}^0(\Omega)$ is the space of piecewise continuous functions on $\Omega$.
\subsection{Composite trapezoidal rule}
In order to obtain the value of the bond at origination, i.e. $V(0,S_0,r_0;T)$, by means of expression (\ref{bondValue}) the computation of an integral term is required. The approximation of this integral is carried out by using a suitable numerical integration procedure. More precisely, we employ the classical composite trapezoidal rule with $M+1$ points, where $M$ is the number of coupons, in the following way:
\begin{equation}
\int_0^T u_2(0,S_0,r_0;\tau_1)d\tau_1\approx \frac{h}{2}\left[u_2(0,S_0,r_0;0)+2\sum_{j=1}^{M-1}u_2(0,S_0,r_0;k_j)+u_2(0,S_0,r_0;T) \right]
\end{equation}
where $h=\frac{T}{M}$, $k_j=jh$ for $j=1,...,M-1$ and $u_2(0,S_0,r_0;0)=r_0$.
\section{Numerical results}\label{results}
In order to show the good performance of the model and numerical methods explained in Section \ref{numericalmethod}, we present some numerical results. In the following examples, the value of some of the parameters involved in the underlying factors are taken from the literature, in particular from \cite{ddp2017} where the authors calibrate the model parameters to market data. More precisely, first the interest rate model is calibrated to zero-coupon bonds (ZCB) and next the model is calibrated to CDS spreads, the price of which is obtained by means of an asymptotic expansion method.

In both examples, the number of elements and nodes of the finite element meshes employed in the numerical solution of the problems are shown in Table \ref{meshes}.

\subsection{Example 1}
First, we consider the simple case of the valuation of  default-free zero-coupon bonds with different maturities. In this setting, the valuation problem is reduced to a one-factor model. The purpose of this example is to compare the value of the bonds we obtain with the market zero-coupon curve. In order to obtain the value of the bonds, we solve the IBVP (\ref{ivp})-(22) with initial condition $g(x_1,x_2)=1$ (corresponding to $h(S,r)=1$)and only taking into account as underlying factor the interest rate. The value of the parameters involved in the interest rate model are the ones collected in Table \ref{dataset1}. The values of the zero-coupon curve and the approximated ones obtained by solving the here proposed model are presented in Table \ref{zcb}. For computing the numerical solution of Table \ref{zcb} we consider Mesh 32 from Table \ref{meshes} and the time step $\Delta \tau=\frac{1}{360}$ (one day).

\begin{table}[!h]
\begin{center}
\begin{tabular}{|c|c|c|}
\hline
&\text{Number of elements}& \text{Number of nodes}\\
\hline
\text{Mesh 4} & 16 & 81\\
\hline
\text{Mesh 8} & 64 & 289\\
\hline
\text{Mesh 16} & 256 & 1089\\
\hline
\text{Mesh 32}& 1024 & 4225\\
\hline
\end{tabular}
\caption{Different finite element meshes (number of elements and nodes).}\label{meshes}
\end{center}
\end{table}

\begin{table}[!h]
\begin{center}
\begin{tabular}{|c|}\hline{Parameters of the defaultable stock price model}\\
\hline
$a_1=0.0337851$ \\
$a_2=0.0523625$\\
$b_1=0.0026639$\\
$b_2=0.0027968$\\
$c=0.0435673$ \\
$\beta=-0.268496$\\
\hline{Parameters of the interest rate model}\\
\hline
$\kappa=0.04520533766268042$\\
$\delta=0.02146900332086033$\\
$\theta=0.10334921942765922$\\
\hline{Correlation coefficient}\\
\hline
$\rho=0.0$\\
\hline
\hline
{Initial conditions}\\
\hline
$S_0=1.0$\\
$r_0=-0.009159871729892612$\\
\hline
\end{tabular}
\caption{Parameters of the model for the UBS bond.}\label{dataset1}
\end{center}
\end{table}

\begin{table}[!h]
\begin{center}
\begin{tabular}{|c|c|c|}
\hline
\text{Maturity (years)}&\text{Market ZCB}& \text{Model ZCB}\\
\hline
1& 1.00229 & 1.006751\\
\hline
2& 1.00372 & 1.009062\\
\hline
3& 1.00333 & 1.007495\\
\hline
4& 1.00099 & 1.002601\\
\hline
5& 0.995825 & 0.994902\\
\hline
6& 0.987805 & 0.984889\\
\hline
7& 0.976833 & 0.973024\\
\hline
8& 0.963223 & 0.959738\\
\hline
9& 0.947687 & 0.945429\\
\hline
10& 0.932845 & 0.930463\\
\hline
\end{tabular}
\caption{Market and model values of the ZCB.}\label{zcb}
\end{center}
\end{table}

\subsection{Example 2}
Next, we consider the valuation of two real defaultable bonds traded in the market and issued by different firms. For this purpose, as in \cite{ddp2017}, we assume that the coefficients $a(t)$ and $b(t)$ in (\ref{lambda}) are linearly dependent on time and their expresions are given by $a(t)=a_1 t +a_2$, $b(t)=b_1 t +b_2$, where $a_1$, $a_2$, $b_1$ and $b_2$ are constants.

On one hand, we take into account the pricing of a bond from UBS with maturity 5 years and a face value of 100. The bond pays annually coupon rates of 1.25 basis points and the recovery rate at the event of default is 40$\%$. The model parameter values for this example are collected in Table
\ref{dataset1}. In this case the correlation coefficient $\rho$ is assumed to be zero. Next, in Table \ref{UBSBond} we present the value of the bond for different meshes and time steps. In this case, we can appreciate that the price of the bond converges to 102.62.

On the other hand, we present a correlated case. More precisely, we address the valuation of a bond from JP Morgan with maturity 5 years and a face value of 100. The bond pays annually coupon rates of 3.25 basis points and the recovery rate at the event of default is again 40$\%$. For this second example the parameter values of the model are the ones which appear in Table \ref{dataset2}. As we have just pointed out the correlation coefficient $\rho$ is different from zero. Finally, the value of this bond is shown in Table \ref{JPMBond}. In this case, the value of the bond converges to 103.57.

In both cases, in order to obtain the value of the bond for we need to solve for each coupon payment date the IBVP (\ref{ivp})-(22) with maturity equal to those dates and with initial condition  $g(x_1,x_2)=1$ (corresponding to $h(S,r)=1$) or $g(x_1,x_2)=\exp(-\kappa T)(x_2-\tilde{x}_2^\infty)$ (corresponding to $h(S,r)=r$. More precisely, in both examples the maturity of the bond is 5 years and the frequency of coupon payments is annually, thus to obtain the value of the bond we need to solve the problem (\ref{ivp})-(22) 10 times, i.e. 5 times with one initial condition to obtain the value of $u_1$ and 5 times with the other initial condition to obtain the value of $u_2$.

\begin{figure}[h!]
\centering
\includegraphics[width=12cm]{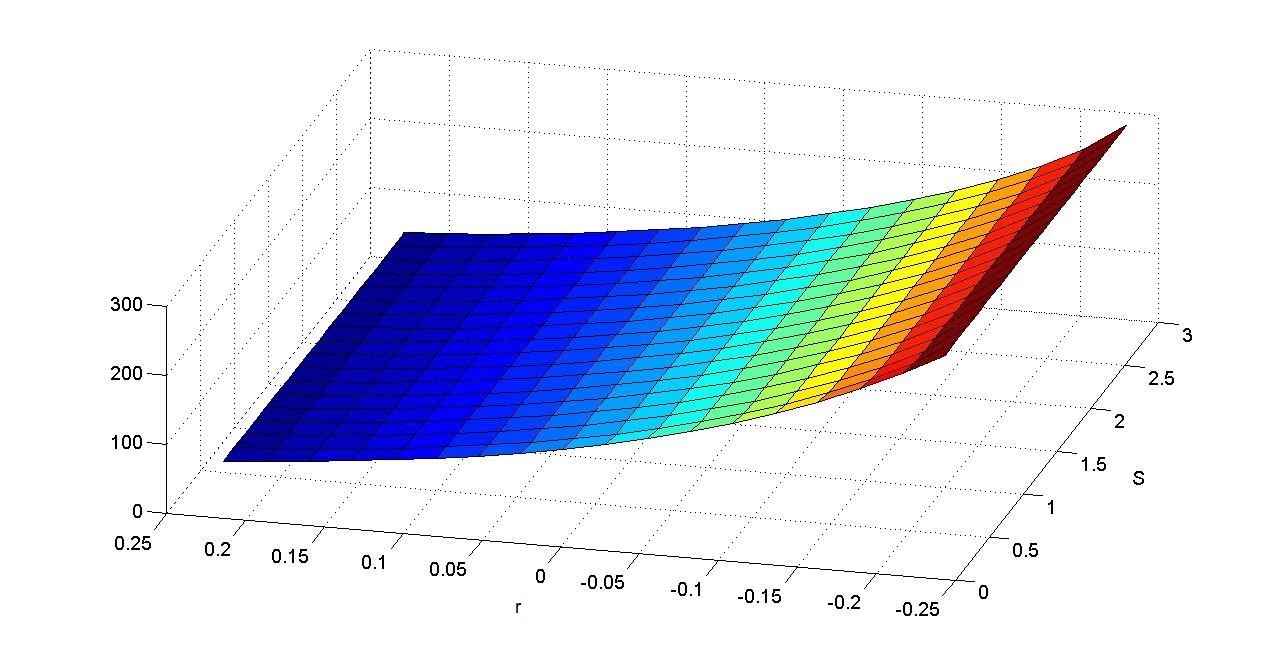}
\caption{Value of the UBS bond}\label{UBSbond}
\end{figure}

\begin{figure}[h!]
\centering
\includegraphics[width=12cm]{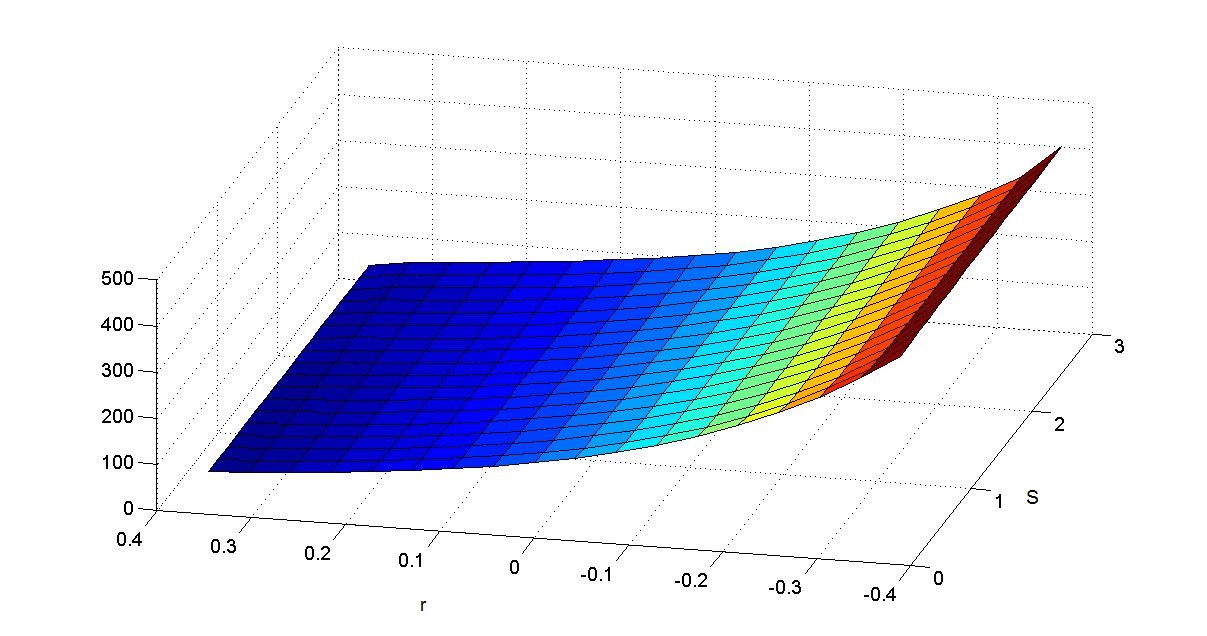}
\caption{Value of the JP Morgan bond}\label{JPMbond}
\end{figure}

Next, in Figures \ref{UBSbond} and \ref{JPMbond} we show the mesh value of the UBS bond and the JP Morgan bond, respectively. Both figures are obtained with the finest mesh and with time step $\Delta \tau=\frac{1}{360}$ (one day).

Finally, for UBS and JPM bonds the Table \ref{PDE_MC} shows a comparison between the results obtained with the proposed numerical method for the PDE model, a crude Monte Carlo technique and the asymptotic approximation method introduced in \cite{MR3323557,MR3276811}. More precisely, for the PDE numerical results we consider Mesh 32 with $360$ time steps while we use $100000$ simulations in Monte Carlo and we show the $95 \%$ confidence interval.

In the fourth column, we present the prices of the UBS and JPM bonds computed with the asymptotic approximation method. This method consists of approximating the solution $u$ of the parabolic PDE \eqref{and1} by applying Theorem 3.2 in \cite{MR3323557}. Note that the obtained value from the PDE numerical methods belongs to the confidence interval and is more accurate than the asymptotic approximation when compared with the real market prices.

Concerning computational time, the asymptotic method takes around 0.047 seconds to obtain one value and Monte Carlo simulation takes 74 seconds for UBS bond and 45 seconds for JPM bond to obtain one value, while the numerical solution of the PDE takes around 440 seconds to obtain the bond values at all the 4225 mesh nodes.

\begin{table}[!h]
\begin{center}
\begin{tabular}{|c|c|c|c|c|}
\hline
\text{Time steps per year}&\text{Mesh 4}& \text{Mesh 8} & \text{Mesh 16} & \text{Mesh 32} \\
\hline
90 & 102.499496 & 102.603837 & 102.616859 & 102.619028 \\
\hline
180 & 102.499599 & 102.605953 & 102.617976 & 102.619681 \\
\hline
360 & 102.500454 & 102.606351 & 102.618421 & 102.620069  \\
\hline
\end{tabular}
\caption{Value of the UBS bond for different meshes and time steps.}\label{UBSBond}
\end{center}
\end{table}

\begin{table}[!h]
\begin{center}
\begin{tabular}{|c|}\hline{Parameters of the defaultable stock price model}\\
\hline
$a_1=0.0312763$ \\
$a_2=0.0356952$\\
$b_1=0.00038362$\\
$b_2=0.00172115$\\
$c=0.346622$ \\
$\beta=-0.223027$\\
\hline{Parameters of the interest rate model}\\
\hline
$\kappa=0.14485883018483803$\\
$\delta=0.01330207057173363$\\
$\theta=0.03467342840511061$\\
\hline{Correlation coefficient}\\
\hline
$\rho=0.497108$\\
\hline
\hline
{Initial conditions}\\
\hline
$S_0=1.0$\\
$r_0=0.01469383913023823$\\
\hline
\end{tabular}
\caption{Parameters of the model for the JP Morgan bond.}\label{dataset2}
\end{center}
\end{table}

\begin{table}[!h]
\begin{center}
\begin{tabular}{|c|c|c|c|c|}
\hline
\text{Time steps per year}&\text{Mesh 4}& \text{Mesh 8} & \text{Mesh 16} & \text{Mesh 32} \\
\hline
90 & 103.725041 & 103.596891 & 103.572191 & 103.570225 \\
\hline
180 & 103.841153 & 103.605155 & 103.575270 & 103.572747 \\
\hline
360 & 103.794567 & 103.602389 & 103.576483 & 103.574147 \\
\hline
\end{tabular}
\caption{Value of the JP Morgan bond for different meshes and time steps.}\label{JPMBond}
\end{center}
\end{table}

\begin{table}[!h]
\begin{center}
\begin{tabular}{|c|c|c|c|}
\hline
 &\text{PDE solution}& \text{MC confidence interval}& \text{asymptotic approximation}\\
\hline
\text{UBS}& 102.620069 & [102.52477413, 102.72696887] & 102.566\\
\hline
\text{JPM}& 103.574147 & [103.55570424, 103.65668368] & 103.584\\
\hline
\end{tabular}
\caption{Comparison of PDE numerical solution with 360 time steps and Mesh 32 with respect to Monte Carlo $95 \%$ confidence interval with 100000 simulations and 2nd-order asymptotic approximation. Note that the real market prices of the bonds of the UBS and JPM are respectively 102.62 and 103.57.}\label{PDE_MC}
\end{center}
\end{table}

\section{Conclusions}\label{conclusions}
In this paper we have considered the valuation of a non callable defaultable coupon bond where the underlying stochastic factors are the interest rate and the defaultable stock price. The pricing problem is posed as a sequence of IBVPs. More precisely, two PDE problems with different initial conditions with maturity each coupon payment date need to be solved. Once the numerical solution of these problems is carried out, the value of the bond is computed by means of an expression which also involves the computation of an integral term.

In order to obtain a numerical solution of the PDE problems, we have proposed appropriate numerical methods based on Lagrange-Galerkin formulations. More precisely, we combine a Crank-Nicolson semi-Lagrangian scheme for time discretization with biquadratic Lagrange finite elements for space discretization. Moreover, the integral term which is involved in the computation of the bond value is approximated by means of the classical composite trapezoidal rule. Finally, we show some numerical results in order to illustrate the behaviour of the proposed methods and its a agreement with the ones obtained by an alternative Monte Carlo technique  and an asymptotic aproximation method.

\section{Acknowledgements}

First and fourth authors have been partially supported by the Spanish government (Ministerio de Econom\'{\i}a y Competitividad, project MTM2016-76497-R) and Xunta de Galicia (grant GRC2014/044, including FEDER funds). Second, third and fourth authors have been supported by EU H2020-MSCA-ITN-2014 (WAKEUPCALL Grant Agreement 643045).

\end{document}